\documentclass[%
 reprint,
superscriptaddress,
 amsmath,amssymb,
 aps,
 prb,
]{revtex4-1}

\usepackage{graphicx}
\usepackage{dcolumn}
\usepackage{bm}
\usepackage{caption,subcaption}
\usepackage{layouts}

\begin{document}

\preprint{APS/123-QED}

\title{
Electronic structure based descriptor for characterizing local atomic environments
}

\author{Jan Jenke}
\affiliation{%
 ICAMS, Ruhr-Universit\"at Bochum, D-44801 Bochum, Germany
}%
\author{Aparna P. A. Subramanyam}
\affiliation{%
 ICAMS, Ruhr-Universit\"at Bochum, D-44801 Bochum, Germany
}%
\author{Marc Densow}
\affiliation{%
 ICAMS, Ruhr-Universit\"at Bochum, D-44801 Bochum, Germany
}%
\author{Thomas Hammerschmidt}
\affiliation{%
 ICAMS, Ruhr-Universit\"at Bochum, D-44801 Bochum, Germany
}%
\author{David G. Pettifor}%
\affiliation{%
 University of Oxford, Oxford, United-Kingdom
}%
\author{ Ralf Drautz}
\affiliation{%
 ICAMS, Ruhr-Universit\"at Bochum, D-44801 Bochum, Germany
}%

\date{\today}

\begin{abstract}
A quantitative descriptor of local atomic environments is often required for the analysis of atomistic data. Descriptors of the local atomic environment ideally provide physically and chemically intuitive insight. This requires descriptors that are low-dimensional representations of the interplay between atomic geometry and electronic bond formation. The moments of the local density of states (DOS) relate the atomic structure to the electronic structure and bond chemistry. This makes it possible to construct electronic structure based descriptors of the local atomic environment that have an immediate relation to the binding energy. We show that a low-dimensional moments-descriptor is sufficient as the lowest moments, calculated from the closest atomic neighborhood, carry the largest contributions to the local bond energy. Here, we construct moments-descriptors that project the space of local atomic environments on a 2-D map.
We discuss in detail the separation of various atomic environments and their connections in the map. The distances in the map may be related to energy differences between local atomic environments as we show by analytic considerations based on analytic bond-order potentials (BOP) and by numerical assessment using TB and density-functional theory calculations. 
Possible applications of the proposed moments-descriptors include the classification of local atomic environments in molecular-dynamic simulations, the selection of structure sets for developing and testing interatomic potentials, as well as the construction of descriptors for machine-learning applications.\\
\\
This article has been published under the copyright of the American Physical Society.
\end{abstract}

\maketitle

\section{\label{sec:introduction}Introduction}
Descriptors are frequently employed in the statistical analysis of physical and chemical properties of materials. For example, local bond-order parameters are used in structure identification \cite{PhysRevB.28.784, doi:10.1063/1.2977970}, parameterizations of interatomic interactions utilize the bispectrum \cite{PhysRevLett.104.136403}, smooth overlaps of atomic positions \cite{PhysRevB.87.184115} or atom-centered symmetry functions \cite{doi:10.1063/1.3553717, PhysRevB.83.153101, PhysRevB.95.014114}, to name just a few. Molecular properties are predicted based on the Coulomb matrix \cite{PhysRevLett.108.058301}, Fourier series of atomic radial distribution functions \cite{QUA:QUA24912} and the bag of bonds method \cite{doi:10.1021/acs.jpclett.5b00831}. Materials properties are evaluated from combinations of atomic quantities \cite{doi:10.1021/acs.chemmater.5b04299, PhysRevLett.114.105503, PhysRevB.85.104104}, partial radial distribution functions \cite{PhysRevB.89.205118} or structural and 
electronic 
fingerprints \cite{doi:10.1021/cm503507h}. 
Descriptors are further used in the classification of structural properties in structure maps \cite{doi:10.1021/acs.chemmater.5b04299} and property maps \cite{PhysRevLett.114.105503, doi:10.1021/cm503507h, PhysRevB.85.104104}. 

Here we show that the moments of the DOS may serve as robust descriptors of the local atomic structure that allow for an intuitive grouping and classification of atomic environments in a map. The local electronic density of states (DOS) intimately relates the energy on the one hand to the atomic structure on the other hand. The formal relation between the moments of the DOS and the local crystal structure was introduced explicitly with the moments theorem \cite{doi:10.1080/00018736700101495}. The moments theorem enables the computation of the moments of the local DOS without the computationally expensive calculation of the eigenspectrum and is used for linear scaling expansions of the band energy \cite{PhysRevLett.73.122, PhysRevB.51.9455, PhysRevLett.63.2480, doi:10.1142/S0217979293000640, PhysRevLett.71.3842, PhysRevB.53.12694, SILVER1996115, PhysRevB.53.12733, PhysRevB.74.174117,PhysRevB.84.214114} and more recently also to define difference vectors between pairs of crystal structures \cite{cryst6020018}. 
Moments-based expansions exploit that in general the lowest moments of the DOS have the largest contribution to the cohesive energy \cite{PhysRevB.74.174117,PhysRevB.84.214114} and therefore are, together with geometrical constraints, critical in the determination of low energy atomic environments. 

In the present paper we exploit the fact that the lowest moments have in general the largest contribution to the energy, which allows us to project the space of atomic environments on a 2-D map. The 2-D map of local atomic environments can be sampled with high-throughput density functional theory (DFT) calculations and may be employed for scanning local atomic environments, for example, for the selection of crystal structures for testing or developing interatomic potentials, the separation or classification of crystal structures, and the comparison of existing descriptors in a low dimensional space.
The paper is organized as follows. In Sec. \ref{sec:moments_descriptors} we introduce the moments of the density of states and discuss how they may serve as descriptors. This allows us to set up a 2-D map of local atomic environments in Sec.~\ref{sec:map}. In Sec. \ref{sec:relation_to_structural_energy} we relate structural energy differences in the map to differences in structural stability obtained by TB and DFT calculations. In Sec. \ref{sec:applications} an outlook on possible applications of the map of local atomic environments and the moments-descriptors is given and in Sec. \ref{sec:conclusion} we conclude our findings. 

\section{\label{sec:moments_descriptors}Moments-descriptors}
In electronic structure calculations, such as DFT or TB the electronic DOS is usually obtained by diagonalizing the Hamiltonian $\hat{H}$.  The Hamiltonian thereby contains the complete information required for characterizing the electronic structure of a material and depends in particular on the positions of the atoms as well as their chemistry. Therefore, the moments of the DOS incorporate information on the atomic structure as well as the chemistry of a material. 
The moments of the DOS are explicitly linked to the crystal structure and chemistry through the moments theorem. We summarize the moments theorem \cite{doi:10.1080/00018736700101495} for a local, orthonormal set of basis functions. The $N$-th moment of the local DOS $n_{inlm}(E)$ of orbital $n$ with angular momentum $l$ and projection $m$ on atom $i$ is defined by
\begin{equation}
\mu_{inlm}^{(N)} = \int^{\infty}_{-\infty} E^N n_{inlm} \left(E\right) \mathrm{d} E\,, \label{eq:moments}
\end{equation}
with the energy $E$. The moments theorem states that the $N$-th moment of the local DOS can be computed by summing over all self-returning paths of Hamiltonian matrix elements of length $N$ that start and end at orbital $inlm$,
\begin{equation}
\begin{split}
\mu_{inlm}^{(N)} = \sum_{\substack{i_1 n_1 l_1 m_1,\\ i_2 n_2 l_2 m_2, \dots}} & \langle inlm | \hat{H} | i_1 n_1 l_1m_1 \rangle \cdot \\
                                                                               & \langle i_1 n_1 l_1 m_1  | \hat{H} | i_2 n_2 l_2 m_2 \rangle \cdot \\
                                                                               & \langle i_2 n_2 l_2 m_2  | \hat{H} | \dots \rangle \cdots \\
                                                                               & \langle \dots |\hat{H} | inlm \rangle\,.
\end{split}
\label{eq:moments_theorm}
\end{equation}

By averaging contributions of different magnetic quantum numbers rotationally invariant atomic moments are obtained,
\begin{equation}
\mu_{inl}^{(N)} = \frac{1}{2l+1} \sum_{m = -l}^{+l} \mu_{inlm}^{(N)}\,.
\end{equation}
The atomic moments are by construction also invariant with respect to reflection, translation of the atomic structure and to permutation of atoms of the same species and fulfill the basic requirements for an atomic scale descriptor \cite{PhysRevB.87.184115, doi:10.1063/1.3553717}.

As the DOS is strictly positive, we may normalize the zeroth moment to one, $\mu_{i nl}^{(0)} = 1$. The first moment corresponds to the center of gravity of the DOS,
\begin{equation}
\mu_{inl}^{(1)}  = E_{inl}\,.
\label{eq:choice_atomic_first_moment}
\end{equation}
By an appropriate shift $E \to E - E_{inl}$ of the energy scale we achieve $\mu_{inl}^{(1)} =0$. The second moment, the root mean square width of the local DOS, is the lowest moment that depends on the atomic environment,
\begin{equation}
\begin{split}
\mu_{i nl}^{(2)} =   \frac{1}{2l+1}  \sum_{m i' n' l'm'} & \langle inlm |\hat{H}| i'n'l'm' \rangle \cdot \\
& \langle i'n'l'm' | \hat{H} | inlm \rangle \,. \label{eq:mu2}
\end{split}
\end{equation}
Through the dependence of the second moment on the Hamiltonian matrix, the second moment depends explicitly on the interatomic distances. As our focus is on the characterization of local atomic environments without an explicit scaling length or density dependence, we need to remove the distance dependence from the second moment. This is achieved by homogeneously scaling interatomic distances such that 
\begin{equation}
\mu_{i nl}^{(2)} = 1 \,.
\label{eq:choice_atomic_second_moment}
\end{equation}
With the above scaling, the third and fourth moment $\mu_{i nl}^{(3)}$ and $\mu_{i nl}^{(4)}$ (that contribute information on the skewness and bimodality of the local DOS) are the lowest two moments that depend explicitly on the local atomic structure.

Instead of working with the third and fourth moment directly, we rewrite the moments in the form of recursion coefficients \cite{HAYDOCK198011}. The recursion coefficients are the matrix elements of a Hamiltonian that is transformed onto a 1-D semi-infinite chain. With the normalization $\mu^{(0)}=1$, $\mu^{(1)}=0$ and $\mu^{(2)}=1$ the corresponding recursion coefficients are given by
\begin{align}
a^{(1)} & = \mu^{(3)} \,, \\
b^{(2)} & = \sqrt{\mu^{(4)} - \left( \mu^{(3)} \right)^2 - 1} \,,
\label{eq:a1_b2}
\end{align}
with the common index $inl$ omitted. The recursion coefficients $a^{(1)}$ and $b^{(2)}$ are the two moments-descriptors that we use to span the map of local atomic environments. The recursion coefficient $a^{(1)}$ measures the skewness of the local DOS, while the recursion coefficient $b^{(2)}$ is a dimensionless shape parameter \cite{Pettifor}, which is smaller than one for a bimodal local DOS and larger than one for a unimodal DOS. 

The recursion coefficients are independent parameters which may in principle attain independently any value for $a^{(1)}$ or any positive value for $b^{(2)}$. This is not true for the moments which are not independent. For example, from $(b^{(2)})^2 \geq 0$ we immediately obtain
\begin{equation}
\mu^{(4)} \geq \left( \mu^{(3)} \right)^2 + 1 \,.
\label{eq:inequality}
\end{equation}
Inequalities for higher moments can also be derived \cite{doi:10.1093/biomet/21.1-4.361}.

For the computation of the moments-descriptors $a^{(1)}$ and $b^{(2)}$ for a particular atomic structure a Hamiltonian $\hat{H}$ is required in the evaluation of Eq.\eqref{eq:moments}. While this Hamiltonian could be taken from DFT, we focus on obtaining a map of local atomic environments that may be used for different materials. We achieve this by using model TB Hamiltonians (Eqs. \ref{eq:d_valent_model}, \ref{eq:sp_valent_model}) that show good transferability across the transition metals and the $sp$-elements, respectively~\cite{PhysRevB.17.1209,  0953-8984-3-5-001}. 
These simple models of the bonding chemistry enable us to analyze the influence of valence character and band filling on the binding energy and the resulting structural stability.
Details of the TB models are summarized in App.~\ref{sec:chemistry}.
In addition, the choice of a TB model allows for an efficient calculation of the moments of the local DOS of atom $i$ by Eq.\eqref{eq:moments_theorm} without computation of the TB eigenspectrum. 
The numerical calculation of moments and recursion coefficients in this work was performed with the BOPfox program \cite{BOPfox}.

\section{Map of local atomic environments \label{sec:map}}
\begin{figure*}
\includegraphics[width=\textwidth]{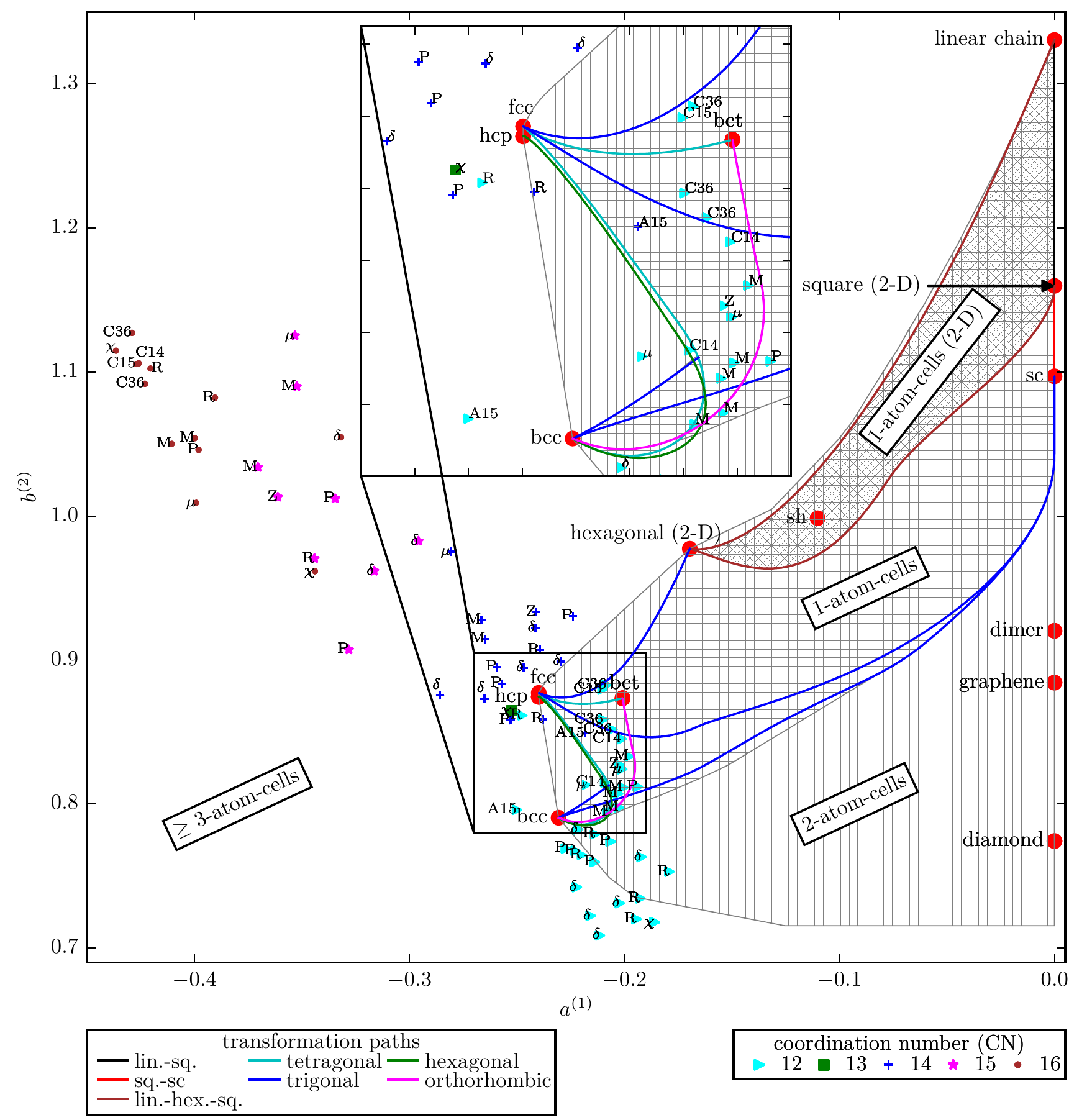}
\caption{Map of local atomic environments for a $d$-valent Hamiltonian. Depicted are different crystal structures, transformation paths and estimated envelopes that demarcate the regions in which crystal structures with one, two or more inequivalent atoms may exist. Red filled circles correspond to crystal structures with only one atomic environment. Further symbols indicate the position of differently coordinated atoms in more complex crystal structures, the common names for TCP phases are given next to the symbols. A square pattern shows the region into which all structures with only one atom in the primitive cell fall. The region of 2-D 1-atom structures is further marked by diagonal lines. The existence region of crystal structures that contain a maximum of two atoms in the primitive cell is indicated by vertical lines. This region includes the region of 1-atom cells. Crystal structures with three or more atoms in the primitive cell can in principle reach any position in the map. Transformation paths 
between different crystal structures are shown using colored lines. The region around the close packed phases (bcc, fcc, hcp) is magnified.
}
\label{fig:first_map}
\end{figure*}

We will next introduce the map of local atomic environments and illustrate its efficiency for separating crystal structures. We will further motivate the descriptors by relating them to the binding energy in Sec. \ref{sec:relation_to_structural_energy}. Fig.~\ref{fig:first_map} shows the $d$-valent map of local atomic environments that is spanned by $a^{(1)}$ and $b^{(2)}$, which are computed with the $d$-valent TB model. The red filled circles correspond to crystal structures with only one atomic environment. Other symbols indicate the differently coordinated atoms in more complex crystal structures. Furthermore, existence regions for structures with one, two or more inequivalent atoms are marked as patterned areas while transformation pathways between different structures are shown as lines.
The envelopes of the existence regions are estimated from the positions of a large set of random structures (cf. Sec.~\ref{sec:filling_the_map}).

\subsection{Simple crystal structures \label{sec:simple_crystal_structures}}
The map of local atomic environments provides a clear separation of simple crystal structures with only one atomic environment (filled red circles). For the linear chain (linear), 2-D square lattice (square 2-D) and the simple cubic (sc) structure, graphene and diamond the third moment is zero. They are therefore placed on the line $a^{(1)} = 0$ on the right of the map. The linear chain, the 2-D square lattice and the simple cubic structure are ordered according to their dimensionality. They are followed by the dimer, graphene and diamond. Dimer, graphene and diamond have characteristically lower values of $b^{(2)}$, which leads to their stabilization in some materials as we will discuss in Sec.~\ref{sec:TB_energy}. 

Towards the left of the map of local atomic environments, we find the close-packed structures face-centered cubic (fcc), hexagonal close-packed (hcp) and body-centered cubic (bcc). The map of local atomic environments places fcc and hcp, which typically have a very similar cohesive energy, almost on top of each other. The values of the moments-descriptors differ only due to their small fourth moment contributions \cite{PhysRevB.74.174117}. The small difference is a result of the different stacking sequence of ABC for fcc and ABA for hcp, which has a small effect on the self-returning paths that reach the third layer. \footnote{A better differentiation between fcc and hcp can easily be set up by using higher recursion coefficients for the axes, such as $a^{(2)}$ and $b^{(3)}$. However, the lowest moment contributions $a^{(1)}$ and $b^{(2)}$ are most important for a general structural differentiation, such that a higher dimensional map would be required for a general structural differentiation\cite{cryst6020018}.}

Among the simple structures, the map places the bcc structure next to fcc and hcp. Once more this is intuitive as the three structures are realized in transition metal elements. 
The special body-centered-tetragonal (bct) \cite{PhysRevB.66.094110} structure is close to the close-packed structures. 
Between the close-packed structures and the open structures with $a^{(1)} = 0$ are the simple hexagonal and body-centered tetragonal structure as well as the 2-D close-packed hexagonal lattice (2-D hexagonal).
 
By construction the positions in the map of local atomic environments may be related to the local density of states of the different crystal structures: The linear chain, the 2-D square lattice, the simple cubic structure, the dimer, graphene and diamond all have a symmetric DOS ($\mu^{(3)}=0$). The dimer, which has a perfect bimodal DOS ($b^{(2)}=0$) for $s$-orbitals shows a finite value of $b^{(2)}$ for $d$-orbitals and the DOS of graphene and diamond are even more bimodal. The 2-D hexagonal lattice has the most skewed DOS among the 2-D structures with one atom in the primitive cell. The bcc structure is more bimodal and less skewed than fcc and hcp \cite{PhysRevB.74.174117}. 

Some of the simple crystal structures are related by structural transformation paths that can readily be included in the map of local atomic environments (cf. Fig. \ref{fig:first_map}). More details on the different transformation paths are compiled in App. \ref{sec:creation_of_trafo_paths}. All transformation paths starting from the close-packed structures initially go to the right in the map of local atomic environments. The two transformation paths lin.-sq. and lin.-hex.-sq. form a closed area. This area is in agreement with the estimated envelope of the 2-D structures with one atom in the primitive cell. The trigonal transformation path connects bcc with the simple cubic structures. The tetragonal transformation path from bcc to fcc is almost on top of the hexagonal transformation from bcc to hcp indicating that the intermediate structures along both paths are similar to each other. After approaching fcc the tetragonal transformation path abruptly changes its direction towards the special bct structure. 
This structure is also 
reached by the orthorhombic transformation \cite{PhysRevB.66.094110}. The orthorhombic path returns from bct to bcc along the same path as from bcc to bct. The trigonal path, the tetragonal path and the lin.-hex. path also form parts of the envelope for 3-D structures with one atom in the primitive cell.

\subsection{\label{sec:inequivalent_sites}Crystal structures with multiple inequivalent lattice sites}
For structures with several inequivalent atomic environments a symbol is displayed in the map for each atomic environment. We show the different atomic environments of topologically close-packed (TCP) phases that are briefly introduced in App.~\ref{sec:tcp_phases}. The moments of the DOS have been applied to quantify the difference between TCP phases and to identify trends of the local moments with coordination number \cite{cryst6020018,t.hammerschmidt20081, PhysRevB.83.224116}. The 12-fold coordinated atoms in the TCP phases are close to the fcc and hcp structures in the map. For atoms with higher coordination the absolute values of $a^{(1)}$ and $b^{(2)}$ increase. The sublattices of the $\chi$-phase also follow this trend, the 13-fold coordinated site is close-by to the hcp structure. We observe a clear trend of coordination in the map of local atomic environments, see Fig. \ref{fig:first_map}: Atoms with similar coordination numbers are close to each other, but 
still can be distinguished in the map of local atomic environments. Atoms with high coordination leave the region of simple structures with one or two atoms in the primitive cells, indicating that these atomic environments can only occur in combination with lattice sites of lower coordination. 

\subsection{\label{sec:filling_the_map}Random structures}
We furthermore use randomly generated structures to evaluate domains in the map of local atomic environments that may be covered by structures with 1 or 2 atoms in the primitive cell. Details on the construction of the random structures are given in App.~\ref{sec:creation_of_random_structures}. The domain of structures with 1-atom cells is surprisingly small. It covers the region from the linear chain and the simple cubic lattice at $a^{(1)} = 0$ to the close-packed bcc and fcc phases. A significant area of the domain corresponds to 2-D structures. For two atoms in the primitive cell, the 1-atom domain is expanded significantly to lower values of $b^{(2)}$ and comprises the dimer, graphene and diamond structures. Crystal structures with three or more atoms in the cell can in principle reach any point in the map. Figure \ref{fig:first_map} also shows that the transformation paths between simple structures provide envelopes of 2-D and 3-D structures with one atom in the primitive cell. 

\section{Electronic structure interpretation of moments-descriptors\label{sec:relation_to_structural_energy}}
\subsection{Relation of moments-descriptors to binding energy}
We rationalize structural stability across the maps of local atomic environments by a TB model. We do not account for charge transfer between atoms or between different orbitals within an atom (i.e. the promotion energy) or magnetism. With these approximations, the energy may be written as \cite{0022-3719-21-1-007,0965-0393-23-7-074004}
\begin{equation}
U = U_{\mathrm{bond}} + U_{\mathrm{rep}} \,.
\end{equation}
To lowest order the repulsive contribution may be assumed to be pairwise \cite{0022-3719-21-1-007,0965-0393-23-7-074004},
\begin{equation}
U_{\mathrm{rep}} = \frac{1}{2} \sum_{ij} \Phi_{ij}(r_{ij}) \,.
\end{equation}
 It is possible to estimate the energy difference between two structures without explicit parameterization of the repulsive contribution to the energy by making use of the structural energy difference theorem \cite{0022-3719-19-3-002}: if two structures are compared at identical repulsive energy, the energy difference between the two structures is given to first order by the difference in bond energy,
\begin{equation}
\Delta U \approx \left[ \Delta U_{\mathrm{bond}} \right]_{\Delta U_{\mathrm{rep}} = 0} \,.
\end{equation}
For computing $\Delta U$ one can in many cases
\footnote{We note that the assumption Eq.\eqref{eq:betasq} has its limitations and is not always valid. For example, the $\chi$-phase structure of Mn is stabilized over close-packed hcp that are taken by the isoelectronic Tc or Re by a softer repulsion with an exponent that is smaller than two as Mn does not have $d$-states in the core \cite{PhysRevB.89.134102}. The same holds for carbon, where graphite is stabilized over the diamond structure by the same mechanism \cite{Pettifor}.}
assume that $U_{\mathrm{rep}}$ is dominated by the overlap repulsion \cite{Pettifor,PhysRevB.83.224116,PhysRevB.72.144105}
\begin{equation}
\Phi_{ij}(r) \propto \beta^2(r) \,, \label{eq:betasq}
\end{equation}
where the distance dependence of $\beta(r)$ is proportional to the distance dependence of the Hamiltonian matrix elements. From Eq.\eqref{eq:mu2} it then follows that the second moment is proportional to the atomic repulsion
\begin{equation}
\mu^{(2)} \propto \sum_{j} \frac{1}{2} \Phi_{ij} \,.
\end{equation}
By requiring that all structures in the map of local atomic environments have identical second moments, Eq. \eqref{eq:choice_atomic_second_moment}, the energy difference between two structures may be estimated from the bond energy difference  $\Delta U_{\mathrm{bond}}$. 

The atomic bond energy may be obtained from integrating the density of states up to the Fermi level $E_{\mathrm{F}}$,
\begin{equation}
 U_{\mathrm{bond}, i} = \int^{E_{\mathrm{F}}} (E - E_i) n_i(E) \,dE \,.
\label{eq:Ubond}
\end{equation}
The analytic bond-order potentials (BOP) provide an expansion of the bond energy in terms of its moments and the Fermi energy \cite{PhysRevB.74.174117,PhysRevB.84.214114},
\begin{equation}
\begin{split}
U_{\mathrm{bond}, i} = & 2(2l+1) b^{(\infty)} \left\lbrace \sum_{m=0}^{n_{\mathrm{max}}} \sigma_i^{(m)} \left[ \hat{\chi}_{_{m+2}}(\phi_{\mathrm{F}}) \right. \right. \\
&\left.\left.  - \gamma_{0}  \hat{\chi}_{_{m+1}}(\phi_{\mathrm{F}}) +  \hat{\chi}_{_{m}}(\phi_{\mathrm{F}})\right] \vphantom{\sum_{m=0}^{n_{\mathrm{max}}}} \right\rbrace \,, \label{eq:BOPUbond}
\end{split}
\end{equation} 
where $b^{(\infty)}$ scales the energy range of the density of states to the interval  $\epsilon=\left[-1,1\right]$. The expansion coefficients are given by the Chebyshev moments of the density of states 
\begin{equation}
\sigma_i^{(m)} = \int_{-1}^1 U_m(\epsilon) n_i(\epsilon)\,d\epsilon\,,\label{eq:sigma}
\end{equation}
with the Chebyshev polynomials of the second kind $U_m$ and may therefore be obtained from the moments of the density of states, cf. Eq.\eqref{eq:moments}. 

The response functions $\hat{\chi}_n$ depend on $\phi_{\mathrm{F}}$, which is defined by the Fermi energy $E_{\mathrm{F}} = a^{(\infty)} + 2 b^{(\infty)}  \cos \phi_{\mathrm{F}}$, 
where $a^{(\infty)}$ is the center of the band, $ \gamma_{0}  = (\mu^{(1)} -  a^{(\infty)})/(2 b^{(\infty)})$ and
\begin{align}
\hat{\chi}_{_{1}}(\phi_{\mathrm{F}}) &=  1-\frac{\phi_{\mathrm{F}}}{\pi}+\frac{1}{2\pi}\sin(2\phi_{\mathrm{F}}) \,, \label{eq:chi1} \\
\hat{\chi}_{_{n}}(\phi_{\mathrm{F}}) &= \frac{1}{\pi} \left[ \frac{ \sin(n+1) \phi_{\mathrm{F}}}{n+1} - \frac{ \sin(n-1) \phi_{\mathrm{F}}}{n-1} \right] \, . \label{eq:chin}
\end{align}
The response function $\hat{\chi}_n$ of order $n$ has $n-2$ nodes in the band. In particular, the third order response function $\hat{\chi}_3$ is positive for less than half full band and negative for more than half full band. The fourth order response function $\hat{\chi}_4$ is negative at the band edges and positive in the band center.

The bond energy Eq.\eqref{eq:BOPUbond} approaches its TB reference value for $n_{\mathrm{max}} \to \infty$. If the expansion is terminated at $n_{\mathrm{max}}=4$, the structural trends across the $sp$-valent elements may still be described \cite{PhysRevB.72.144105}. Furthermore, the difference between the bcc and fcc or hcp structure is to lowest order given by the fourth moment, while resolving the much smaller energy difference between fcc and hcp requires six moments ($n_{\mathrm{max}}=6 $) \cite{Pettifor,PhysRevB.74.174117}. Higher moments are mainly required for a quantitative match of the TB reference energy and in practice most BOP calculations are performed with $n_{\mathrm{max}}=9 $.

The expansion of the bond energy Eq.\eqref{eq:BOPUbond} may be applied to discuss trends in crystal structure stability. When an expansion coefficient $\sigma^{(n)}$ is negative, a positive value of the response function $\hat{\chi}_{_{m+2}}(\phi_{\mathrm{F}}) - \gamma_{0}  \hat{\chi}_{_{m+1}}(\phi_{\mathrm{F}}) +  \hat{\chi}_{_{m}}(\phi_{\mathrm{F}})$ will lower the energy and vice versa. For making contact with the map of local atomic environments we take the simplest possible fourth moment expansion with $n_{\mathrm{max}}=4$, $a^{(\infty)} = a^{(0)} = 0$ and $b^{(\infty)} = b^{(1)} = 1$. Then $\sigma^{(1)} = \sigma^{(2)}=0$, $\sigma^{(3)} = a^{(1)}$ and $\sigma^{(4)} = ( a^{(1)})^2 +( b^{(2)})^2-1$.

At less than half full band the simple metals take the close-packed structures bcc, hcp and fcc. These are stabilized over competing structures by large negative values of $a^{(1)}$and small values of $b^{(2)}$. The details of the ordering from bcc Na over hcp Mg and fcc Al cannot be resolved within the map of local atomic environments as one cannot expect the simple, nearly-free electron metals to be described well within a simple TB approximation. At half full band the response function $\hat{\chi}_3$ is zero while $\hat{\chi}_4$ is at its maximum and therefore a small value of $b^{(1)}$ is favorable and helps to stabilize the diamond structure. The subtle competition between graphite and the diamond lattice in carbon is not covered by this argument as the comparison of the two structures at identical second moment is not adequate\cite{Note2, Pettifor, PhysRevB.89.134102}. Still, graphene is close to diamond in the $d$- and $sp$-map. The dimer, which is stabilized for hydrogen with its half full $s$-orbitals, takes the 
minimum of $b^{(2)}=0$ in an $s$-valent map (not shown).

The transition metals all take close-packed structures, broadly due to the attraction provided by the $s$-electrons, while the $d$-electrons determine the details of the crystal structure. In a map that only takes into account the $d$-valence we may therefore not expect to find the transition metal structures at extreme boundaries of the map. Still, the map places them at large absolute values of $a^{(1)}$ and small values of $b^{(2)}$. As expected the bcc structure, which is stabilized by the response function $\hat{\chi}_4$ at the center of the $d$-band, has a smaller value of $b^{(2)}$ than fcc or hcp, while hcp and fcc shows a slightly more negative value for $a^{(1)}$.

The discussion of the stability of the TCP phases is more involved and has been discussed in detail in \cite{PhysRevB.83.224116, 1367-2630-15-11-115016}. The TCP phases are stabilized by a combination of average band filling and atomic size mismatch. The two factors are of different relevance in the different TCP phases. As the atoms in the different coordination polyhedra have different second moments, a direct discussion of the stability of the TCP phases based on the map of local atomic environments alone is not possible. We note that the TCP phases show small values of $b^{(2)}$ for the 12-fold coordinated sites, some of them even smaller than bcc, while the sites with higher coordination show large negative values of $a^{(1)}$. 

For evaluating the difference in the bond energy between two structures one needs to take into account that the Fermi level of the two structures will in general be different. A first order expansion of the bond energy difference between two structures with the same number of valence electrons $N_e$  at identical first and second moment leads to \cite{PhysRevB.83.224116, Turchi1987}
\begin{equation}
\Delta U_{\mathrm{bond}} = 2(2l+1) b^{(\infty)} \sum_{m=3}^{n_{\mathrm{max}}}  \Delta \sigma^{(m)} \hat{\hat{\chi}}_{_{m}}(\phi_{\mathrm{F}}) \,, \label{eq:DeltaUbond}
\end{equation}
where $ \Delta \sigma^{(m)} $ corresponds to the difference in the expansion coefficients and
\begin{equation}
\begin{split}
\hat{\hat{\chi}}_{_{m}}(\phi_{\mathrm{F}}) = & \frac{1}{\pi}  \left[  \frac{ 2\sin(m+1) \phi_{\mathrm{F}}}{m(m+2)} \right. \\
& \left. - \frac{ \sin(m+3) \phi_{\mathrm{F}}}{(m+2)(m+3)}  - \frac{ \sin(m-1) \phi_{\mathrm{F}}}{m(m-1)} \right]
\end{split}
\label{eq:chichi}
\end{equation} 
and $\phi_{\mathrm{F}}$ depends on the number of valence electrons $N_e$.

For ${n_{\mathrm{max}}} = 4$ we can now approximate the difference in energy between two structures as
\begin{align}
\Delta U_{\mathrm{bond}} = & 2(2l+1)\bigg[  \hat{\hat{\chi}}_{_{3}}(\phi_{\mathrm{F}}) \Delta a^{(1)} \nonumber \\
& +   \hat{\hat{\chi}}_{_{4}}(\phi_{\mathrm{F}}) \left( \Delta (b^{(2)})^2 +  \Delta (a^{(1)})^2  \right)  \bigg] \,, \label{eq:DeltaUbond4}
\end{align}
where we estimate $b^{(\infty)} = b^{(1)} = 1$ and $a^{(\infty)} = a^{(1)} = 0$ as before. We see that the difference between two structures in the map of local atomic environments is approximated by a contribution $\Delta a^{(1)}$ that corresponds to the distance between the structures projected on the $x$-axis and a second contribution that corresponds to the square of the distance between two structures in the map of local atomic environments. The relevance of the two contributions for the energy difference is determined by the number of valence electrons through the response functions Eq.\eqref{eq:chichi}. Independent of the detailed number of valence electrons this implies in general that we may expect that the energy difference between pairs of structures increases with the distance of the structures in the map.

\subsection{Trends of structural stability from TB\label{sec:TB_energy}}
\begin{figure*}
\begin{subfigure}[b]{0.9\columnwidth}
\includegraphics[width=\textwidth]{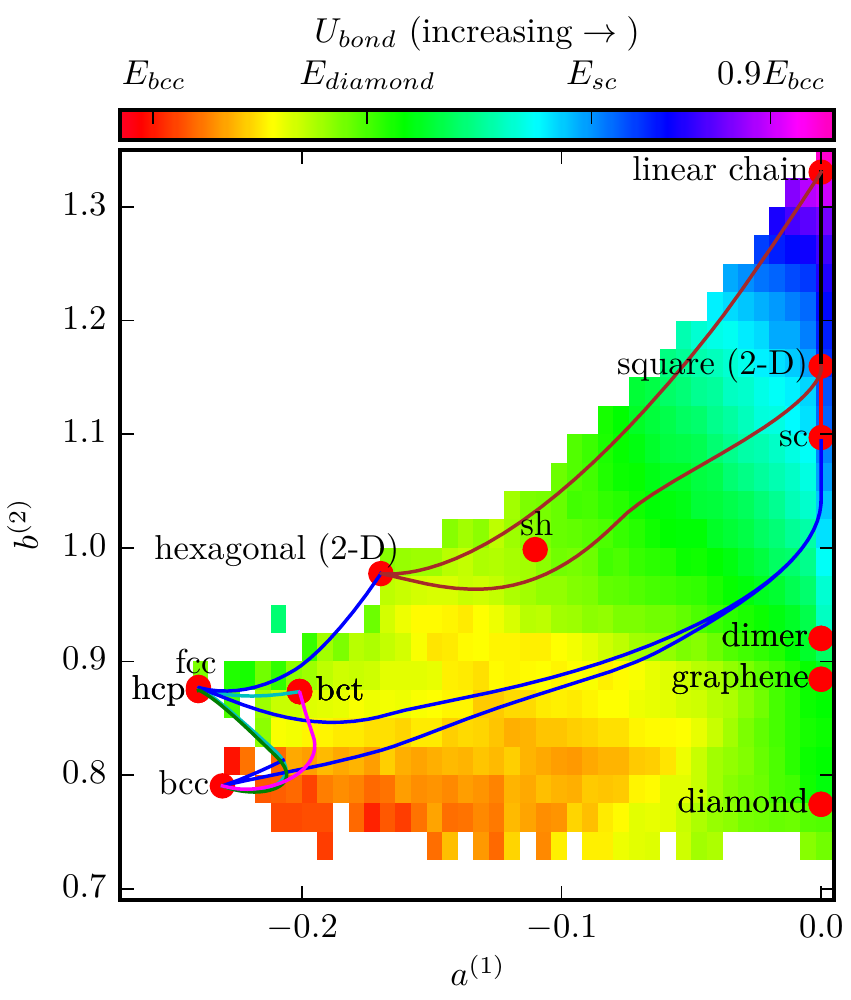}
\caption{Bond energy}
\label{fig:map_TB_energies}
\end{subfigure} \qquad
\begin{subfigure}[b]{0.9\columnwidth}
\includegraphics[width=\textwidth]{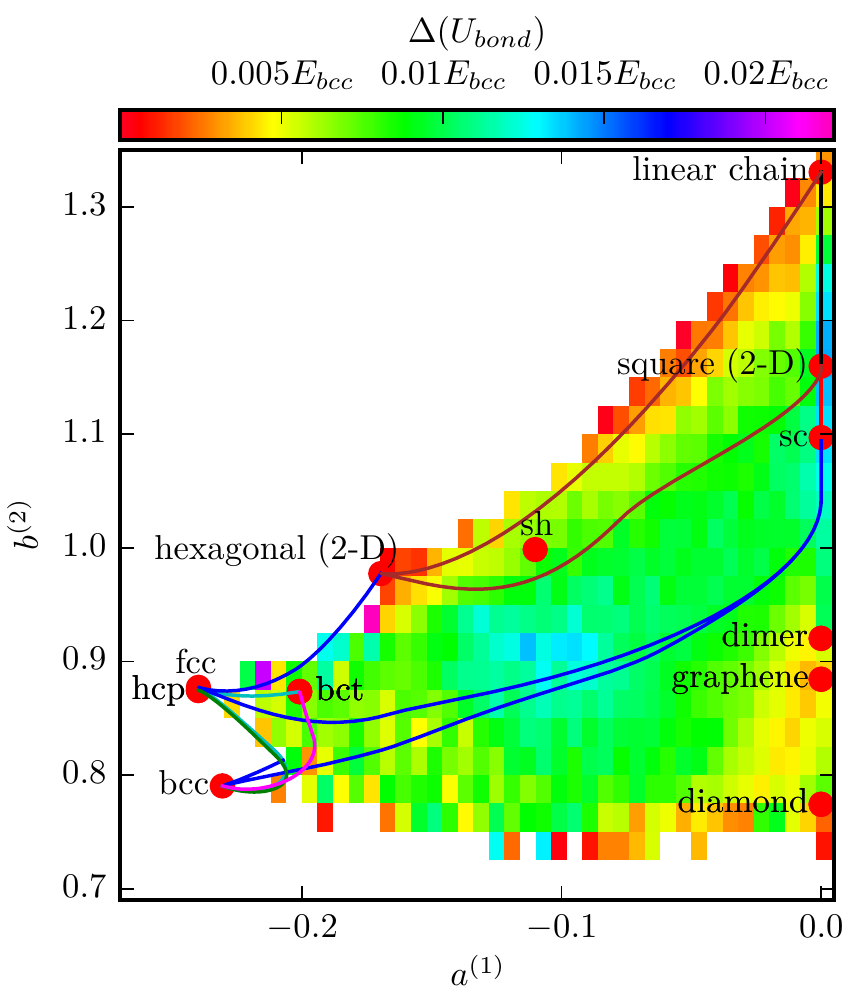}
\caption{Standard error of bond energy}
\label{fig:map_TB_energies_error}
\end{subfigure}
\caption{Analysis of bond energy (Eq.\ref{eq:Ubond}) of 2-atom random structures from canonical $d$-valent TB model evaluated for a band filling of $N_e = 4$ in the $d$-valent map of local atomic environments. The left figure shows the averaged bond energy and the right figure the related standard error that is calculated from different random structures at the same position of the map. Lines correspond to transformation paths introduced in Fig. \ref{fig:first_map}.}
\end{figure*}
We evaluate the bond energy within the TB approximation for the set of random structures in Fig. \ref{fig:prob_dist_d} by numerical calculations with the BOPfox program~\cite{BOPfox}. We choose a canonical $d$-valent TB model (Eq.\ref{eq:d_valent_model}) with a band filling of four, which is close to the maximum bcc stability \cite{PhysRevB.74.174117, PhysRevB.83.224116}. The locally averaged bond energy is shown in Fig. \ref{fig:map_TB_energies}. We observe a smooth increase of the bond energy from bcc, which has the smallest bond energy among all structures, to the linear chain. The overall trend in the map validates our result from Sec. \ref{sec:relation_to_structural_energy} that energy differences between two structures increase with their distance in the map.
The standard error of the bond energy was obtained from many different random structures that are projected to a given location in the map of local atomic environments and is displayed in Fig. \ref{fig:map_TB_energies_error}. The standard error is in the order of $1\%$ of the cohesive energy and hence much lower than the range of energy values in the map of local atomic environments. In other words, the two moments-descriptors allow for the separation of crystal structures that have an energy difference that is greater than about $1\%$ of the cohesive energy. This demonstrates that the descriptors of the map of local atomic environments are excellent predictors for structural 
stability as a direct consequence of the relation between geometric environment and electronic structure provided by the moments theorem, Eq. \eqref{eq:moments_theorm}. This makes our moments-descriptors distinctly different from purely geometrical descriptors.

\begin{figure*}
\begin{subfigure}[b]{0.32\textwidth}
\includegraphics[width=\textwidth]{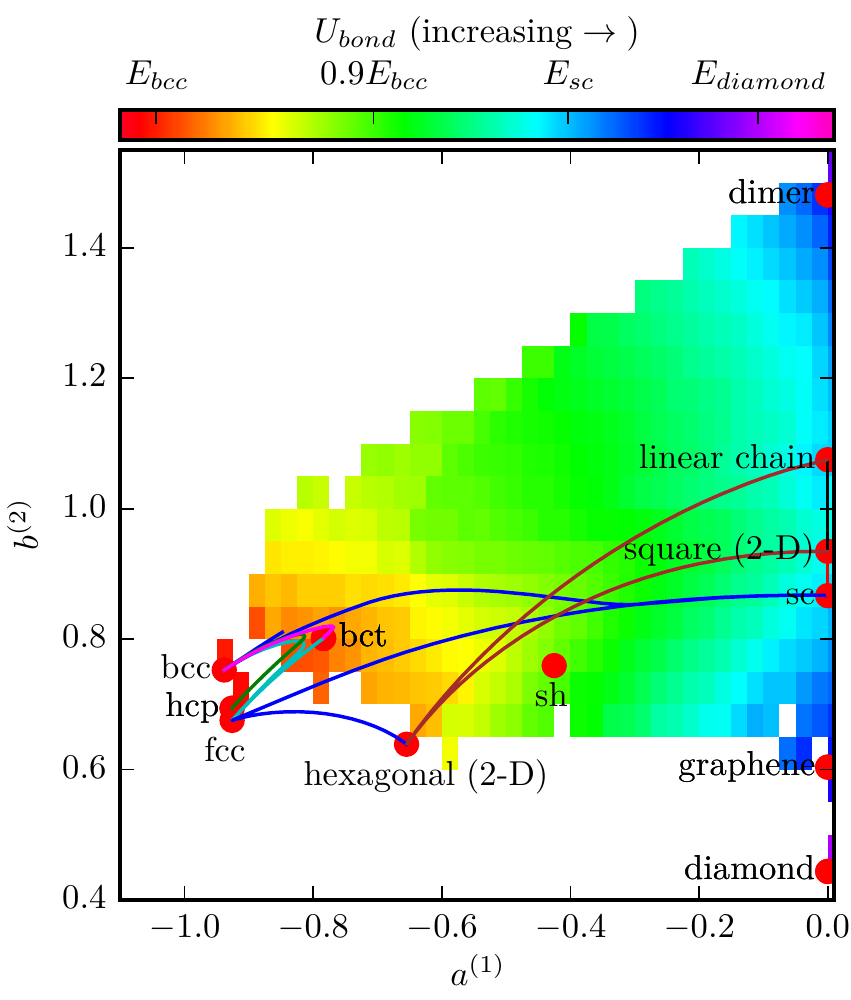}
\caption{$N_e=2$}
\label{fig:ubond_sp_Ne_2}
\end{subfigure}
\begin{subfigure}[b]{0.32\textwidth}
\includegraphics[width=\textwidth]{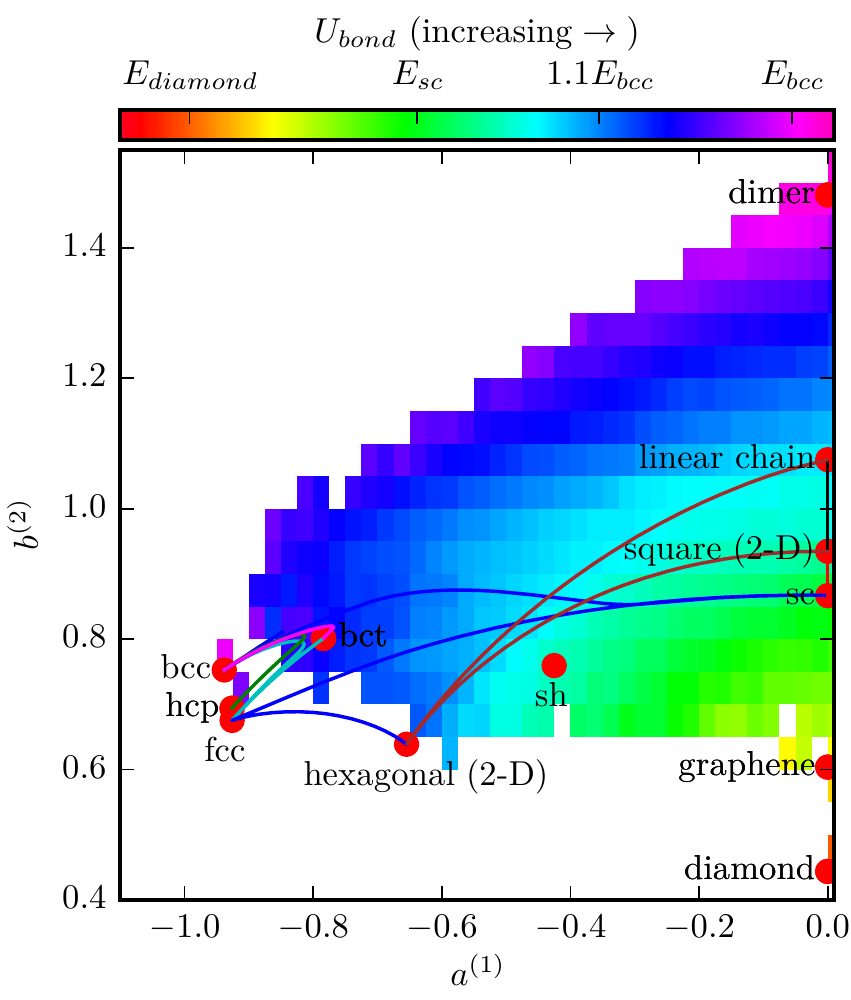}
\caption{$N_e=4$}
\label{fig:ubond_sp_Ne_4}
\end{subfigure}
\begin{subfigure}[b]{0.32\textwidth}
\includegraphics[width=\textwidth]{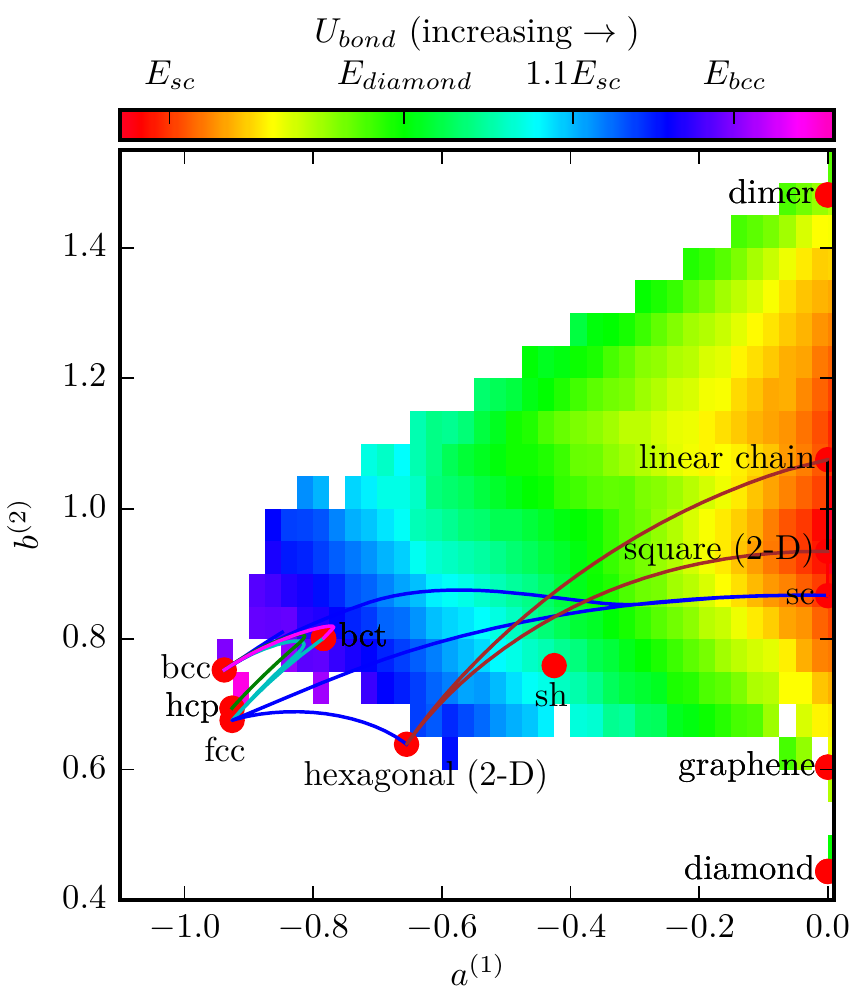}
\caption{$N_e=6$}
\label{fig:ubond_sp_Ne_6}
\end{subfigure}
\caption{Bond energy (Eq.\ref{eq:Ubond}) in the $sp$-valent map of local atomic environments as obtained from the canonical $sp$-valent TB model for different band fillings. Lines correspond to transformation paths introduced in Fig. \ref{fig:first_map}.}
\label{fig:sp_map}
\end{figure*}

We show the bond energy for an $sp$-valent TB model (Eq.\ref{eq:sp_valent_model}) with different band fillings in Fig. \ref{fig:sp_map}. The values for the bond energy were obtained by averaging over many random structures at each position in the map. The values of the atomic recursion coefficients differ from those obtained for the $d$-valent TB model, however, many features of the $d$-valent map of local atomic environments are still present in the $sp$-valent map. As the dimer configuration may be a stable configuration for $sp$-elements \cite{0953-8984-3-5-001}, it is an important feature of the $sp$-valent map that it positions the dimer apart from the other crystal structures. For all band fillings we obtain smooth energy surfaces. As expected at half full band the diamond structure has the lowest bond energy. For low band fillings the stability is shifted towards the close packed phases, at higher band fillings more open structures are favored \cite{PhysRevB.72.144105}. 

\subsection{Trends of structural stability from DFT\label{sec:DFT_energy}}
\begin{figure*}
\begin{subfigure}[b]{0.32\textwidth}
\includegraphics[width=\textwidth]{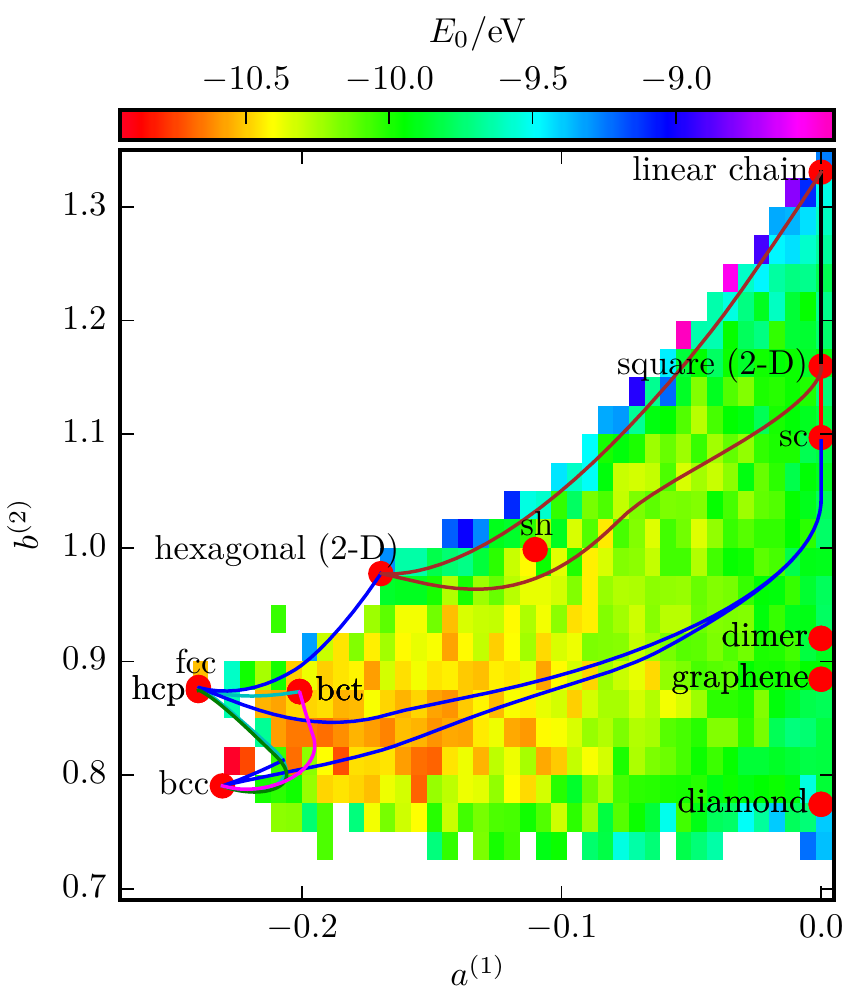}
\caption{Equilibrium energy per atom}
\label{fig:DFT_E0}
\end{subfigure}
\begin{subfigure}[b]{0.32\textwidth}
\includegraphics[width=\textwidth]{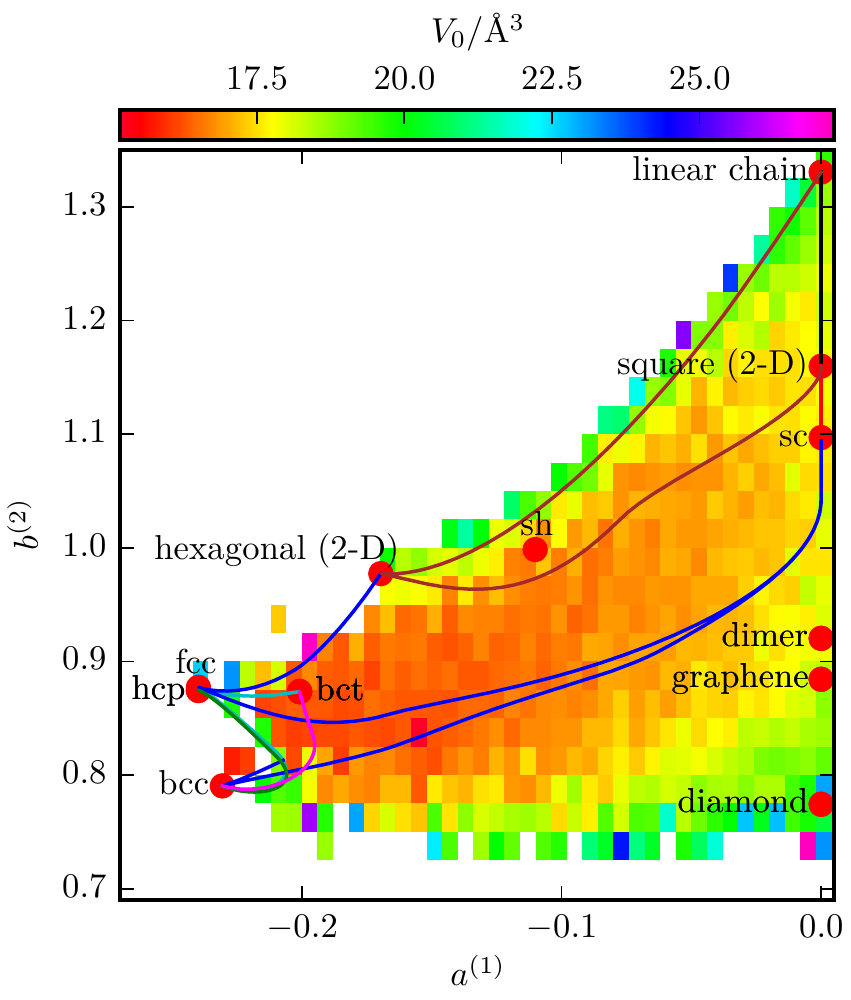}
\caption{Equilibrium volume per atom}
\label{fig:DFT_V0}
\end{subfigure}
\begin{subfigure}[b]{0.32\textwidth}
\includegraphics[width=\textwidth]{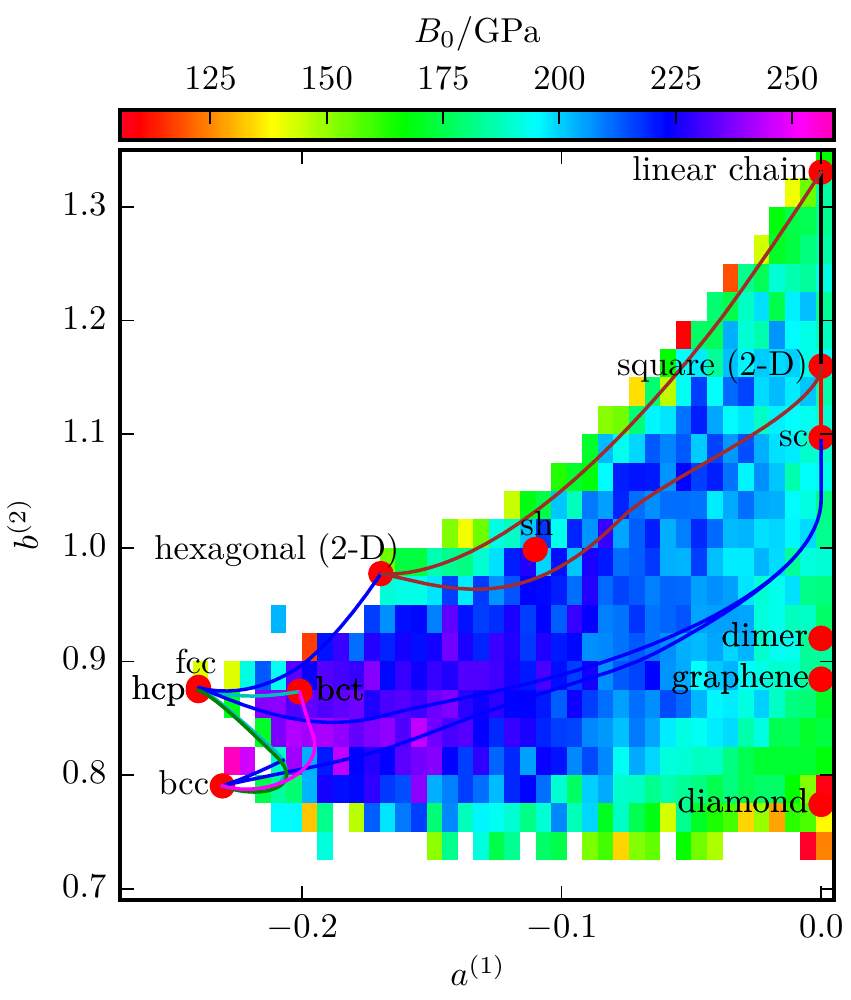}
\caption{Bulk modulus}
\label{fig:DFT_B0}
\end{subfigure}
\caption{DFT calculations of equilibrium energy (left), volume (middle) and bulk modulus (right) for a set of random structures with two atomic sites, both occupied by Mo atoms in the $d$-valent map of local atomic environments. Lines correspond to transformation paths introduced in Fig. \ref{fig:first_map}. The trend of equilibrium energy for Mo obtained by DFT (left) is captured by the bond energy of the corresponding TB calculations at $N_e=4$ (Fig.~\ref{fig:map_TB_energies}).}
\label{fig:DFT}
\end{figure*}

The 4$d$ and 5$d$ transition metals Mo and W may be described using a band filling of approximately four $d$-valence electrons, cf. the TB calculations in Sec.~\ref{sec:TB_energy} \cite{PhysRevB.83.224116,0953-8984-26-19-195501}.
To compare to the TB predictions we performed DFT calculations for Mo for 2-atom random structures with both atomic positions occupied by Mo atoms. Here an evaluation of the approximately 90000 structures that we evaluated for TB was computationally too demanding and we selected a subset of the random structures using the following strategy: The atomic volume of each atom with normalized second moment $\mu^{(2)} = 1$ may be interpreted as a measure for the homogeneity of its atomic surrounding, where a small normalized atomic volume indicates a homogeneous atomic surrounding with equidistant bond lengths. We select from our set of 2-atom random structures a subset of 521 structures with small normalized volume that homogeneously covers the existence region of the 2-atom cells. For these structures we calculated the DFT equilibrium volume, energy and bulk modulus with fixed cell shape and atomic positions. We performed spin-polarized DFT calculations using the VASP software\cite{Kresse-93,Kresse-96-1,Kresse-96-2} with the projector augmented-wave 
method (PAW)~\cite{Bloechl-94} with fourteen valence electrons (Mo\_sv) for molybdenum and employ the generalized gradient approximations (GGA)~\cite{Perdew-96} to the exchange correlation potential. The calculations were performed with a plane-wave cutoff energy of 500\,eV and Monkhorst-Pack~\cite{Monkhorst-76} $\bf k$-point meshes with linear density not more than 0.1\,\AA$^{-1}$. The equilibrium energy $E_0$, volume $V_0$ and bulk modulus $B_0$ were then obtained by fitting energy volume curves with volume scalings of $\pm$10\% to the Birch-Murnaghan equation of state. The results are shown in Figs. \ref{fig:DFT}. 
We observe that the overall trend of the DFT equilibrium energy for Mo (Fig.~\ref{fig:DFT_E0}) is qualitatively captured by the bond energy of the corresponding TB calculations at $N_e=4$ (Fig.~\ref{fig:map_TB_energies}). The lowest energy is taken by bcc and the energy increases with distance in the map of local atomic environments. The bcc structure also takes the smallest equilibrium volume. The bulk modulus is smooth across the map of local atomic environments and largest for bcc. The scattering of the results, in particular at the envelops of the existence region, is an artifact of the relatively smaller number of random structures that we employed in our DFT calculations (see Fig. \ref{fig:prob_dist_d}).

\section{\label{sec:applications}Outlook}
The moments-descriptors use the moments theorem to provide a direct link between the local atomic structure and the local electronic structure that determines the binding energy.
This link is maintained in a 2-D descriptor space, as we demonstrate analytically for the BOPs and numerically for TB and DFT calculations.
Therefore, we expect that our map of local atomic environments will prove useful in applications that relate atomic structure and binding energy.

One potential application is the classification of individual atoms in atomistic simulations like, e.g., molecular dynamics. 
The typically very large number of atoms in such simulations hinders manual analysis and requires tools for automated identification of processes like nucleation, phase transformation, or dislocation movement. 
The computation of the coordinates in the map of local atomic environments provides a straight-forward approach to identify atoms with similar atomic environment and to assign atoms to a particular crystal structure. The representation with low moments ensures that this computation is feasible also in large-scale simulations.

A second example of a potential application is the development and assessment of empirical or semi-empirical interatomic potentials. 
These potentials are typically developed to describe the binding energy of a set of reference structures but often exhibit limited transferability to other structures. The challenges are therefore (i) the identification of reference structures in the development of the potential and (ii) the anticipation of transferability of the interatomic potential. Both aspects can be addressed by the map of local atomic environments as it projects the space of atomic environments on a 2-D space, which enables an extensive and homogeneous sampling of atomic environments.

A third potential application is to use the moments-descriptors for a set of atomic environments directly as features for machine-learning to predict, e.g., the DFT formation energy from the atomic structure. The descriptors incorporate domain knowledge of interatomic bond formation, as demonstrated by the smooth  DFT data Fig.~\ref{fig:DFT} and capture about 99 \% of the cohesive energy of our TB calculations (cf. Sec.~\ref{sec:TB_energy}).

\section{\label{sec:conclusion}Conclusion}
We introduce moments-descriptors for local atomic environments based on the local electronic density of states. The moments depend on the local atomic environment and determine the bond chemistry. We use the lowest two structure dependent moments of the electronic density of states as obtained from canonical $sp$- and $d$-valent TB models to span a 2-D map of local atomic environments.We employ the map of local atomic environments for the discussion of crystal structures. We show that structures with one or two atoms in the primitive cell are bound to specific regions of the map. By making use of the analytic BOP expansion we argue that the lowest energy structure for a specific material should be found close to the boundaries of these regions. We further show that the energy difference between two structures depends on the distance between the structures in the map and carry out extensive TB and DFT calculations to demonstrate this numerically.

For structures with several inequivalent lattice sites the map places similar local environments in close proximity, such that the map of local atomic environments may be employed to sample systematically local atomic environments by projecting the local atomic environments to a low dimensional space.
We point out possible applications of this feature of the moments-descriptors for the classification of local atomic environments in molecular-dynamic simulations, for the selection of structure sets for developing and testing interatomic potentials, as well as for the construction of descriptors for machine-learning applications.

\begin{acknowledgments}
We wish to dedicate this paper to the memory of our coauthor Professor David Pettifor CBE FRS, who sadly passed away before the work was completed. A.P.A.S., T.H., and R.D. acknowledge financial support by the German Research Foundation (DFG) through project C1 of the collaborative research center SFB/TR 103. R.D. acknowledges financial support by the German Research Foundation (DFG) through project C2 of the collaborative research center SFB/TR 103. The authors acknowledge Mike Finnis for helpful discussions.
\end{acknowledgments}

\appendix
\section{\label{sec:chemistry}Description of the chemistry}
In this paper we restrict our analysis to two different TB models, namely a canonical pure $d$-model \cite{PhysRevB.17.1209},
\begin{equation}
  \left.
  \begin{aligned}
    &dd\sigma \\
    &dd\pi \\
    &dd\delta
  \end{aligned} 
  \right\rbrace =
  \left.
  \begin{aligned}
    -&6 \\
    &4 \\
    -&1
  \end{aligned}
  \right\rbrace \beta(r),
  \label{eq:d_valent_model}
\end{equation}
and a $sp$-model \cite{0953-8984-3-5-001} based on Harrison's parametrization \cite{Harrison},
\begin{equation}
  \left.
  \begin{aligned}
    &ss\sigma \\
    &sp\sigma \\
    &pp \sigma \\
    &pp \pi 
  \end{aligned} 
  \right\rbrace =
  \left.
  \begin{aligned}
    -&1.00 \\
    &1.31 \\
    &2.31 \\
    -&0.76
  \end{aligned}
  \right\rbrace \beta(r)\,,
\label{eq:sp_valent_model}
\end{equation}
with
\begin{equation}
\beta(r) = c/r^{5}\,,
\end{equation}
where $c$ is a constant. The pure $d$-model \cite{cryst6020018, PhysRevB.74.174117, PhysRevB.83.224116, PhysRevB.84.155119, PhysRevB.83.184119, 0953-8984-23-27-276004, t.hammerschmidt20081} is often sufficient to describe the elements of the $d$-block and captures structural trends across the $4d$ and $5d$ transition metal series \cite{cryst6020018, PhysRevB.74.174117, PhysRevB.83.224116, t.hammerschmidt20081}. Our $sp$-model disregards the splitting of the onsite elements \cite{PhysRevB.72.144105}. 

We smoothly force the bond integrals to zero at $r=r_{\mathrm{cut}}$ by multiplying with the cutoff function
\begin{equation}
f_{\mathrm{cut}}\left( r \right) = \frac{1}{2} \left( \cos \left( \pi \left[ \frac{r - \left( r_{\mathrm{cut}} - d_{\mathrm{cut}} \right)}{d_{\mathrm{cut}}} \right] \right) + 1 \right)\,,
\end{equation}
where $d_{\mathrm{cut}}$ determines the width of the cutoff function. In our calculations we choose constant values of $r_{\mathrm{cut}}$ and $d_{\mathrm{cut}}$ which include second nearest neighbors within the cutoff sphere for the bcc structure, where the second nearest neighbor distance is close to the first nearest neighbor distance, and just exclude second nearest neighbors for the simple cubic structure. This is achieved by choosing $r_{\mathrm{cut}} \approx 1.25 r_{nn, fcc}$ and $d_{\mathrm{cut}} \approx 0.13 r_{nn, fcc}$, where $r_{nn, fcc}$ is the nearest neighbor distance of fcc with a normalized second moment according to Eq.\eqref{eq:choice_atomic_second_moment}. Note that the exclusion of second nearest neighbors results in a zero third moment for the simple cubic structure but not for the fcc and hcp structure as can be seen in Fig. \ref{fig:first_map}.

\section{\label{sec:creation_of_trafo_paths}Description of transformation paths}
\begin{figure}
\begin{subfigure}[b]{.2\textwidth}
\includegraphics[width=\textwidth]{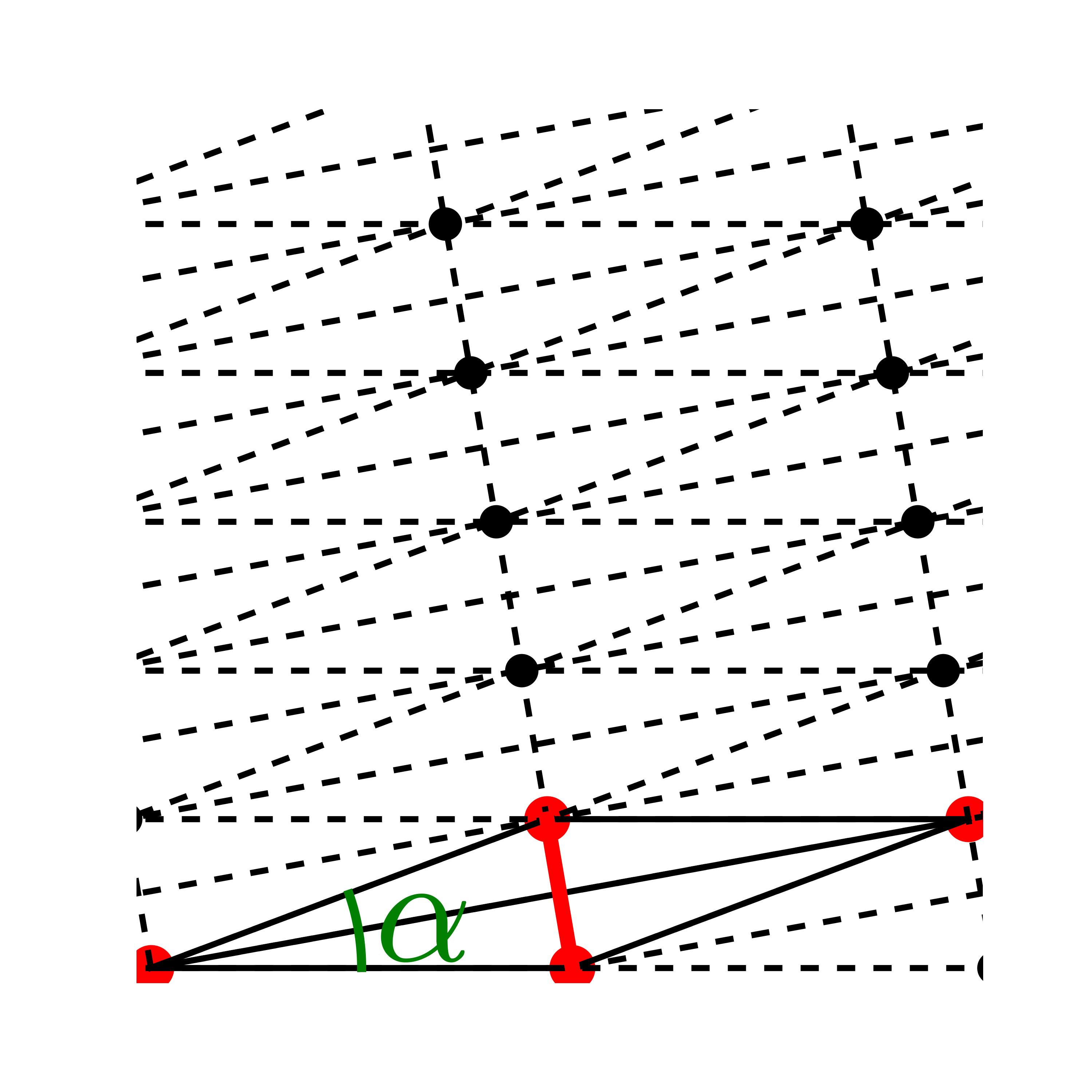}
\caption{$\alpha = \frac{1}{9} \pi$}
\end{subfigure}\qquad
\begin{subfigure}[b]{.2\textwidth}
\includegraphics[width=\textwidth]{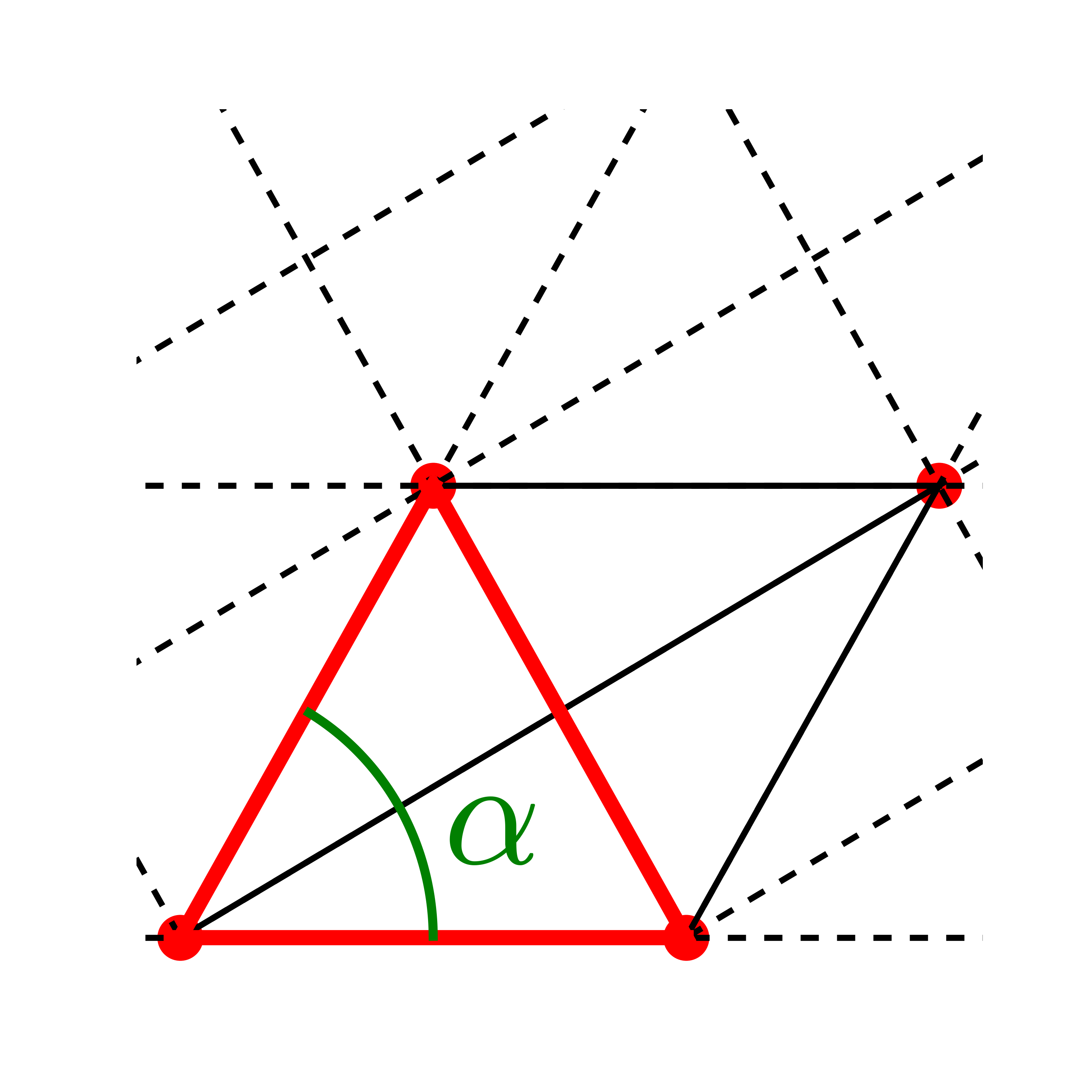}
\caption{$\alpha = \frac{1}{3} \pi$}
\end{subfigure}  
\begin{subfigure}[b]{.2\textwidth}
\includegraphics[width=\textwidth]{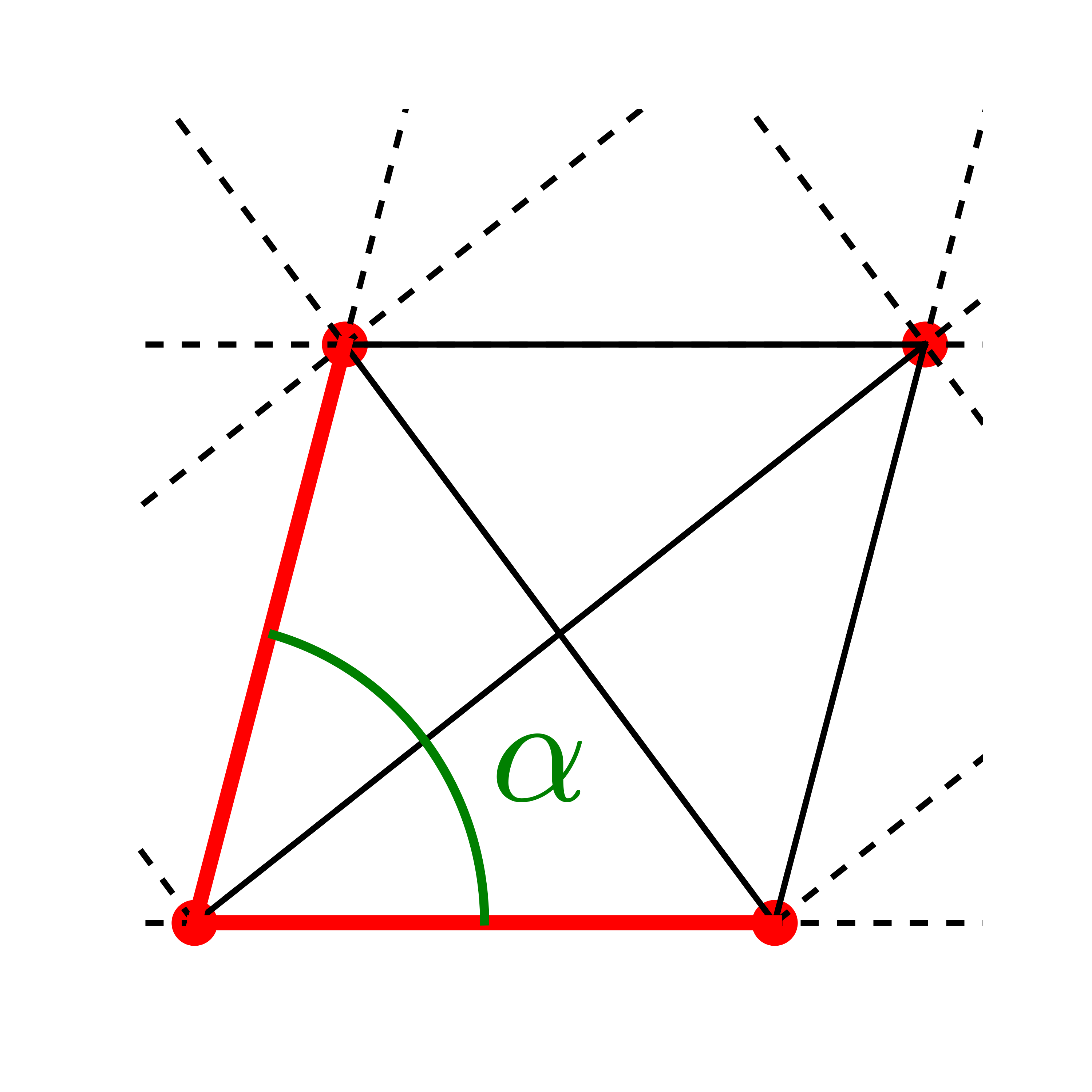}
\caption{$\alpha = \frac{5}{12} \pi$}
\end{subfigure}\qquad
\begin{subfigure}[b]{.2\textwidth}
\includegraphics[width=\textwidth]{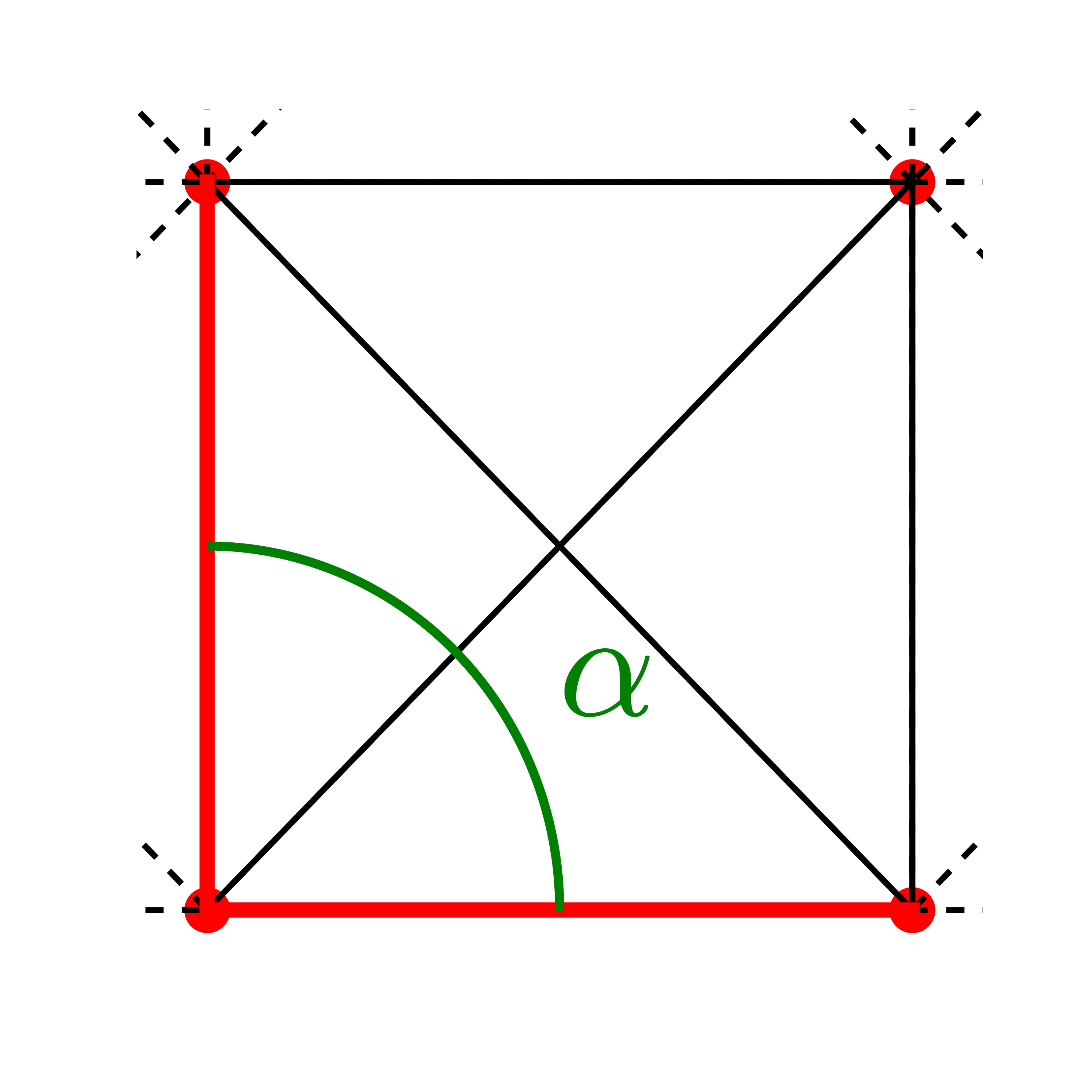}
\caption{$\alpha = \frac{1}{2} \pi$}
\end{subfigure}
\caption{Transformation path connecting the linear chain with the square lattice over the hexagonal lattice. The shortest bond length is marked in red.}
\label{fig:illustration_tranf_path}
\end{figure}  

Transformation paths are continuous deformations of one crystal structure into another. All transformation paths presented in this paper are described by one parameter $p$ changing one structure continuously into the other.
Here, we chose transformation paths that are commonly used to test interatomic potentials (tetragonal, orthorhombic, trigonal, hexagonal)~\cite{0953-8984-26-19-195501, y.s.lin20161, 0965-0393-7-3-306}, as well as transformation paths (lin.-hex.-sq., lin.-sq., sq.-sc) that we found to correspond to envelops of the map of local atomic environments.
The tetragonal transformation path, also called Bain path \cite{bain1924}, connects bcc with fcc. On further continuation it connects fcc with the special body centered tetragonal (bct) structure. This is done by elongating the bcc cell in $[0 0 1]$ direction and compressing it in $[1 0 0]$ and $[0 1 0]$ directions to keep the volume fixed. 

The primitive cell along the path is given by
\begin{align*}
\begin{split}
\mathbf{a_1} = & a (4p)^{-1/3} \begin{pmatrix}-1 & 1 & p\end{pmatrix}^T\\
\mathbf{a_2} = & a (4p)^{-1/3} \begin{pmatrix}1 & -1 & p\end{pmatrix}^T\\
\mathbf{a_3} = & a (4p)^{-1/3} \begin{pmatrix}1 & 1 & -p\end{pmatrix}^T\,, 
\end{split}
\end{align*}
and the atom is located at
\begin{align*}
\mathbf{p_1} = \begin{pmatrix}0 & 0 & 0\end{pmatrix}
\end{align*}
for all values of $p$.
bcc is taken for $p=1$, fcc for $p=\sqrt{2}$ and bct for $p=2^{3/4}$. \\
The orthorhombic transformation path connects bcc with the same special bct structure, which is reached by the Bain path \cite{PhysRevB.66.094110}. Further continuation of the orthorhombic transformation paths leads back to bcc. This is also achieved by an elongation in $[0 0 1]$ direction, however simultaneously a compression in $[1 1 0]$ direction is applied. The primitive cell vectors are therefore 
\begin{align*}
\begin{split}
\mathbf{a_1} = & 4^{-1/3}a \begin{pmatrix}-1 & 1 & p^{1/2}\end{pmatrix}^T\\
\mathbf{a_2} = & 4^{-1/3}a \begin{pmatrix}1 & -1 & p^{1/2}\end{pmatrix}^T\\
\mathbf{a_3} = & 4^{-1/3}a \begin{pmatrix}p^{-1/2} & p^{-1/2} & -p^{1/2}\end{pmatrix}^T
\end{split}
\end{align*}
and the atom is again located at 
\begin{align*}
\mathbf{p_1} = \begin{pmatrix}0 & 0 & 0\end{pmatrix}
\end{align*}
for all values of $p$. bcc is taken for $p=1$, the special bct structure for $p=\sqrt{2}$ and again bcc for $p=2$. 

The trigonal transformation path connects bcc over sc with fcc. Further continuation of the trigonal transformation path, connects fcc with the 2-D hexagonal lattice. This is obtained by an elongation in $[1 1 1]$ direction and a compression in perpendicular directions to keep the volume fixed, the primitive cell is given by
\begin{align*}
\begin{split}
\mathbf{a_1} = & f \begin{pmatrix}p-3 & p+2 & p+2\end{pmatrix}^T\\
\mathbf{a_2} = & f \begin{pmatrix}p+2 & p-3 & p+2\end{pmatrix}^T\\
\mathbf{a_3} = & f \begin{pmatrix}p+2 & p+2 & p-3\end{pmatrix}^T\,,
\end{split}
\end{align*}
with $f = a(25(3p+1))^{-1/3}$. The atom remains at the origin again,
\begin{align*}
\mathbf{p_1} = \begin{pmatrix}0 & 0 & 0\end{pmatrix}\,.
\end{align*}
bcc is taken for $p=1$, sc for $p=2$, fcc for $p=4$ and the 2-D hexagonal lattice for $p \to \infty$.

The bcc to hcp transformation cannot be obtained by a simple deformation of the cell. However, the atoms also have to change their relative positions \cite{0965-0393-7-3-306, PhysRevB.77.174117}. The hexagonal transformation path deforms bcc simultaneously in $[\bar{1} 1 0]$, $[1 1 0]$ and $[001]$ direction. The cell vectors are explicitly given by
\begin{align*}
\begin{split}
\mathbf{a_1} = & 2^{-1/3}a f_1 \begin{pmatrix}-1 & 1 & 0\end{pmatrix}^T\\
\mathbf{a_2} = & 2^{-1/3}a f_2 \begin{pmatrix}0 & 0 & 1\end{pmatrix}^T\\
\mathbf{a_3} = & 2^{-1/3}a \left( f_1 f_2 \right)^{-1} \begin{pmatrix}1 & 1 & 0\end{pmatrix}^T
\end{split}
\end{align*} 
with 
\begin{align*}
\begin{split}
f_1 & = 1 + \alpha_1 (1-p) \\
f_2 & = 1 + \alpha_2 (1-p).
\end{split}
\end{align*}
and 
\begin{align*}
\begin{split}
\alpha_1 & = \left(1-2^{1/6} \sqrt{1.5}\right) / \left(\sqrt{2} - 1 \right) \\
\alpha_2 & = \left(1-2^{1/6}\right) / \left(\sqrt{2} - 1 \right)\,.
\end{split}
\end{align*}
Together with this deformation alternate $(110)$ planes have to be shuffled in $\pm [\bar{1} 1 0]$ direction. We follow the choice of Ref. \onlinecite{0965-0393-7-3-306} and choose
\begin{align*}
s = \frac{2^{-1/6}(p-1)}{4\sqrt{6}(\sqrt{2}-1)f_1}
\end{align*}
as the shuffling factor. The atomic positions in the direct coordinate system are given by
\begin{align*}
\begin{split}
\mathbf{p_1} & = \begin{pmatrix}s & 0 & 0\end{pmatrix}\\
\mathbf{p_2} & = \begin{pmatrix}0.5 + s & 0.5 & 0\end{pmatrix}\\
\mathbf{p_3} & = \begin{pmatrix}-s & 0.5 & 0.5\end{pmatrix}\\
\mathbf{p_4} & = \begin{pmatrix}0.5 -s & 0 & 0.5\end{pmatrix}\,.
\end{split}
\end{align*}
bcc is taken for $p=1$ and hcp for $p=\sqrt{2}$.

The linear chain can be connected with the square lattice over the hexagonal lattice as illustrated in Fig. \ref{fig:illustration_tranf_path}. (lin.-hex.-sq.). The cell vectors of the 2-D cell are given by
\begin{align*}
\begin{split}
\mathbf{a_1} = & a \cos(p)^{-1/2} \begin{pmatrix}1 & 0\end{pmatrix}^T\\
\mathbf{a_2} = & a \cos(p)^{-1/2} \begin{pmatrix}\cos(p) & \sin(p)\end{pmatrix}^T
\end{split}
\end{align*}
The atom remains at position
\begin{align*}
\mathbf{p_1} = \begin{pmatrix}0 & 0 \end{pmatrix}
\end{align*}
for all values of $p$. The square lattice is taken for $p = \pi / 2$, the 2-D hexagonal lattice for $p = \pi/3$ and the linear chain for $p \rightarrow 0$. \\
The linear chain can also be directly connected with the 2-D square lattice by bringing linear chains from infinite separations together until the linear chains are separated by a distance equal to the nearest neighbor distance of the linear chain. (lin.-sq.)\\
Similarly the 2-D square lattice can be connected to the simple-cubic structure by bringing square lattices from infinite separations together until the square lattices are separated by a distance equal to the nearest neighbor distance of the square lattice. (sq.-sc)

\begin{figure*}
\begin{subfigure}[b]{0.9\columnwidth}
\includegraphics[width=\textwidth]{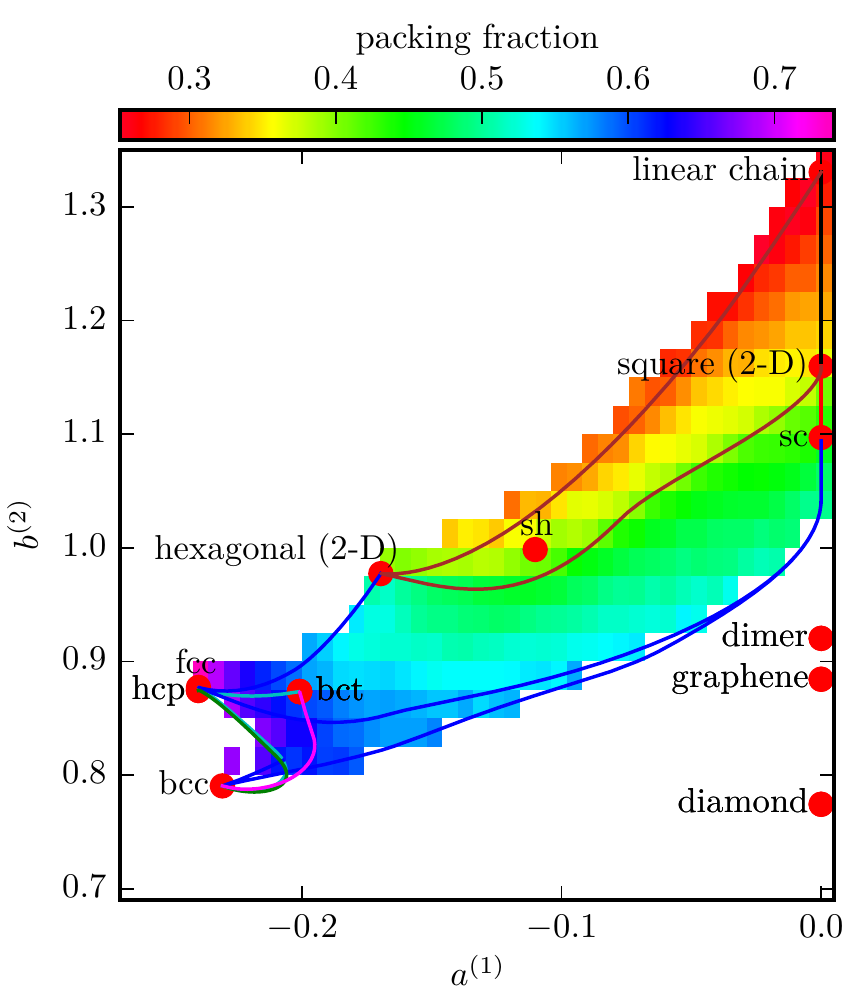}
\caption{Average packing fraction of 1-atom random structures}
\label{fig:map_filled}
\end{subfigure}\qquad
\begin{subfigure}[b]{0.9\columnwidth}
\includegraphics[width=\textwidth]{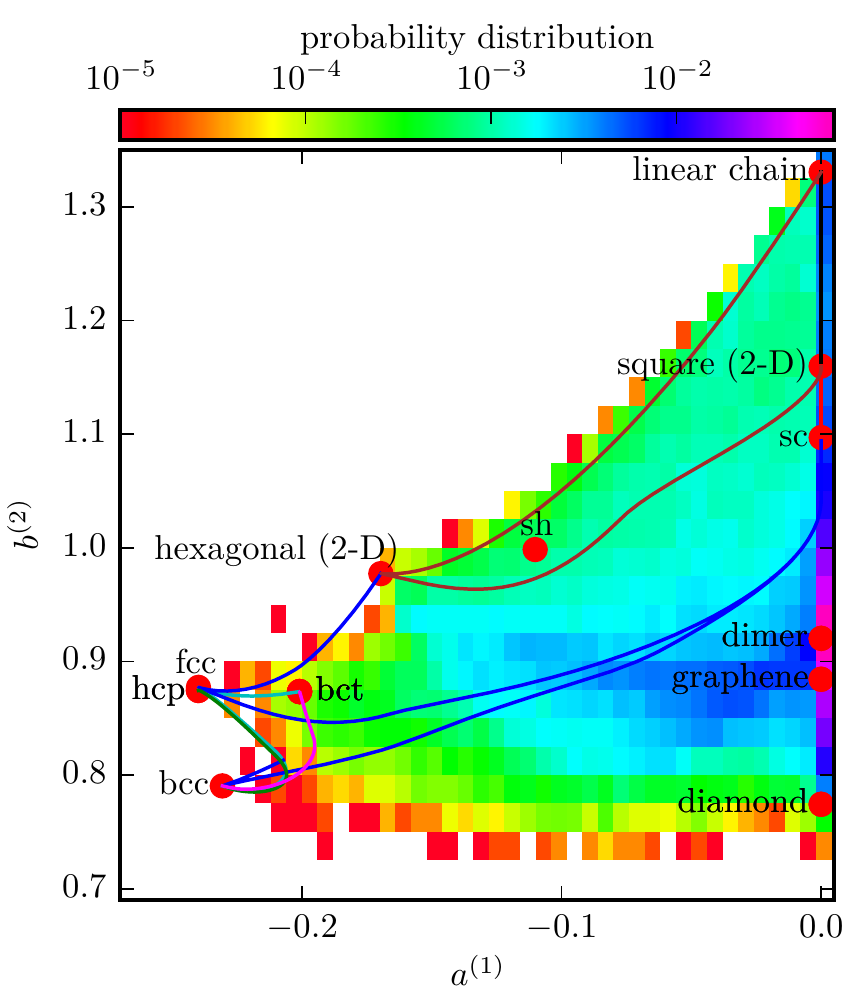}
\caption{Probability distribution of 2-atom random structures}
\label{fig:prob_dist_d}
\end{subfigure}
\caption{Random structures with one or two atoms in the primitive cell in a $d$-valent map of local atomic environments. The left figure shows the average packing fraction that is obtained for the random structures with one atom in the primitive cell. The right figure shows the probability distribution of the random structures with two atoms in the primitive cell when the structures are generated according to the algorithm outlined in App.~\ref{sec:creation_of_random_structures}. Lines correspond to transformation paths introduced in Fig. \ref{fig:first_map}.}
\end{figure*} 

\section{\label{sec:tcp_phases}Topologically close-packed phases}
\begin{table}
\begin{ruledtabular}
\begin{tabular}{cccccccc}
&CN12 &CN13 &CN14 &CN15 &CN16 &$\langle$CN$\rangle$\\
\hline
$\chi$    &12 &12 &- &- &1,4 &13.10\\
C14       &2, 6 &- &- &- &4 &13.33\\
C15       &4 &- &- &- &2 &13.33\\
C36       &4, 6, 6 &- &- &- &4, 4 &13.33\\
$\mu$     &1,6 &- &2 &2 &2 &13.38\\
M         &4, 4, 4, 8, 8 &- &4, 4 &4, 4 &4, 4 &13.38\\
R         &1, 2, 6, 6, 6, 6&- &6, 6 &6 &2, 6 &13.40\\
$\delta$  &4, 4, 4, 4, 4, 4&- &4, 4, 4, 4, 4&4, 4 &4 &13.43\\
P         &4, 4, 4, 4, 8&- &4, 4, 4, 8&4, 4 &4 &13.43\\
Z         &3 &- &2 &2 &- &13.43\\
A15       &2 &- &6 &- &- &13.50\\
\end{tabular}
\end{ruledtabular}
\caption{\label{tab:tcp_phases}Selection of common TCP phases and $\chi$-phase ordered by increasing average coordination number. For each coordination number a list with the number of inequivalent Wyckoff sites is provided.}
\end{table}
Topologically close-packed (TCP) phases consist of coordination polyhedra, which have only triangular faces. The atoms in the TCP phases have coordination numbers 12, 14, 15 or 16. For a selection of common TCP phases the number of atoms with inequivalent Wyckoff positions are listed in Tab.~\ref{tab:tcp_phases}. As in previous works~\cite{PhysRevB.83.224116, 1367-2630-15-11-115016, cryst6020018}, we included the $\chi$-phase in the comparison although it is not a regular TCP phase in the crystallographic sense due to atoms with coordination number 13.

\section{\label{sec:creation_of_random_structures}Random structures}
With random structures we refer to structures generated by randomly choosing their primitive cell and their atomic positions in the primitive cell. The primitive cell is described by the lattice vectors $\textbf{a} = a \textbf{e}_a$, $\textbf{b} = b \textbf{e}_b$, $\textbf{c} = c \textbf{e}_c$. The angle between $\textbf{b}$ and $\textbf{c}$ is named $\alpha$, the angle between $\textbf{a}$ and $\textbf{c}$ is named $\beta$ and the angle between $\textbf{a}$ and $\textbf{b}$ is named $\gamma$. \\
The structure generation is done as follows:
\begin{enumerate}
\item Randomly generate three values $a \leq b \leq c$, with $b/a \leq 3$ and $c/a \leq 3$.
\item Randomly generate angles $\alpha$, $\beta$, $\gamma$ in a range between $0$ and $\pi$ under the condition that the volume is larger than zero\cite{Foadi:au5114}.
\item Place the first atom at the origin and place further atoms randomly in the primitive cell.
\item Even though we generate primitive cells with a finite volume, the generated structure may be effectively 2-D due the finite number of bonds which we obtain due to the choice of the cutoff $r_{\mathrm{cut}}$ of our bond integrals $\beta$. We exclude those structures.
\end{enumerate}
In Fig. \ref{fig:map_filled} we characterize the set of 1-atom random structures in terms of the averaged packing fraction and observe a smooth trend across the map of local atomic environments. The packing fraction is lowest for those 3-D structures which are close to the linear chain in the map. The averaged packing fraction increases towards the bottom and the left of the map of local atomic environments and takes its maximum value close to fcc and hcp, which have the highest possible value \cite{hales2005proof}.

The random structures do not cover the map of local atomic environments homogeneously. In Fig. \ref{fig:prob_dist_d} we show the probability for generating a structure in a particular location of the map of local atomic environments with our algorithm. The probability distribution was obtained from a set of approximately 90000 random structures with two atoms in the primitive cell. It is significantly more likely to generate an open structure with $a^{(1)}=0$ than a close-packed structure close to fcc and hcp.
It can be seen that with this method it is very unlikely to generate structures which are similar to highly symmetric structures. However, it ensures that we do not bias the random structures towards any reference structures.  

\nocite{*}

\bibliography{mybibfile}

\end{document}